\RequirePackage{rotating}

\documentclass[11pt,oneside]{article}

\usepackage{natbib}
\bibpunct[, ]{(}{)}{,}{a}{}{,}%
\def\bibfont{\small}%
\usepackage[hyphens]{url}
\usepackage{hyperref}
\usepackage[hyphenbreaks]{breakurl}
\usepackage{graphicx}
\usepackage{rotating}
\usepackage{graphicx}
\usepackage{booktabs}
\usepackage{amsmath}
\usepackage{amssymb}
\usepackage{tikz}
\usetikzlibrary{shapes}
\usetikzlibrary{patterns}
\usetikzlibrary{arrows,positioning,decorations.pathmorphing,trees}
\usetikzlibrary{arrows.meta}
\usepackage{tikz-cd}
\usepackage{mathtools}
\usepackage{stackengine}
\usepackage{pgfplots}
\usepackage{float}
\usepackage{braket}
\usepackage{eurosym}
\usepackage{hyperref}
\usepackage{comment}
\usepackage{subcaption}
\usepackage{ifthen}
\usepackage{caption}
\usepackage{appendix}
\usepackage[format=default,font=footnotesize,labelfont=bf]{caption}
\usepackage{listings} 
\usepackage{color}
\usepackage{algpseudocode}
\usepackage{algorithm} 
\usepackage{float}
\usepackage{xurl}
\usepackage{array}

\newcommand{\rad}{\text{rad}}

\newcommand{\ch}{\overline{conv}}
\newcommand{\conch}{\overline{conc}}
\newcommand{\conv}{\text{conv}}

\usepackage{fullpage}

\newtheorem{theorem}{Theorem}[section]

\newtheorem{corollary}[theorem]{Corollary}

\newtheorem{lemma}[theorem]{Lemma}

\newtheorem{proposition}[theorem]{Proposition}

\newtheorem{example}{Example}
\newboolean{includeHidden}
\setboolean{includeHidden}{false}
\newcommand{\extR}{\overline{\mathbb{R}}}

\newcommand{\hide}[1]{\ifthenelse{\boolean{includeHidden}}{{\tiny\textbf{HIDDEN:~}#1}}{}}
\newcommand{\eqname}[1]{\tag{#1}}

\newboolean{tablesInText}
\setboolean{tablesInText}{true}
\newboolean{figuresInText}
\setboolean{figuresInText}{true}
\newboolean{footnotesInBrackets}
\setboolean{footnotesInBrackets}{true}

\newtheorem{result}{Result}
\usepackage{lipsum}

\title{Solving Large-Scale Electricity Market Pricing Problems in Polynomial Time}
\author{Mete \c{S}eref Ahunbay \and Martin Bichler \and Teodora Dobos \and Johannes Knörr\thanks{Technical University of Munich, School of Computation, Information and Technology, Department of Computer Science. E-mail: \texttt{mete.ahunbay/bichler/dobos/knoerr@cit.tum.de}}}

\begin{document}

\maketitle

\begin{abstract}
Electricity market operators worldwide use mixed-integer linear programming to solve the allocation problem in wholesale electricity markets. Prices are typically determined based on the duals of relaxed versions of this optimization problem. The resulting outcomes are efficient, but market operators must pay out-of-market uplifts to some market participants and incur a considerable budget deficit that was criticized by regulators. As the share of renewables increases, the number of market participants will grow, leading to larger optimization problems and runtime issues. At the same time, non-convexities will continue to matter e.g., due to ramping constraints of the generators required to address the variability of renewables or non-convex curtailment costs. We draw on recent theoretical advances in the approximation of competitive equilibrium to compute allocations and prices in electricity markets using convex optimization. The proposed mechanism promises approximate efficiency, no budget deficit, and computational tractability. We present experimental results for this new mechanism in the context of electricity markets, and compare the runtimes, the average efficiency loss of the method, and the uplifts paid with standard pricing rules. We find that the computations with the new algorithm are considerably fast for relevant problem sizes. In general, the computational advantages come at the cost of efficiency losses and a price markup for the demand side. Interestingly, both are small with realistic problem instances. Importantly, the market operator does not incur a budget deficit and the uplifts paid to market participants are significantly lower compared to standard pricing rules.
\end{abstract}

\section{Introduction} \label{sec:intro}

Electricity spot markets provide a mechanism for matching supply and demand and for determining the price of electricity at any given moment. The size of electricity spot markets is considerable. For example, the European day-ahead electricity market accounts for more than 98\% of the total electricity consumption in the EU, with a volume of more than 1,530 TWh/year or 200 million Euros average daily value of matched trades \citep{sdac}. 
This market enables trading of electricity for delivery on the following day and is divided into multiple regions or bidding zones. Market participants from each region submit their orders for electricity generation or consumption to their respective Nominated Electricity Market Operator (NEMO). 
Each NEMO aggregates the orders received from its market participants and sends this information to a central clearing platform, known as the Price Coupling of Regions (PCR) solution.
Based on large-scale mixed-integer quadratic  programming problems, the PCR solution calculates the optimal allocation of cross-border transmission capacity, considering the available capacity provided by Transmission System Operators (TSOs) and the aggregated supply and demand information received from the NEMOs. 
The PCR solution also computes the Market Clearing Price (MCP) and the allocated volumes for each bidding zone. This process takes into account the transmission capacity constraints between the zones, ensuring that power flows respect these limits. The PCR solution then sends the results back to the NEMOs, which in turn notify the market participants of their scheduled generation or consumption, as well as the clearing prices. Finally, the TSOs implement the resulting power flows across borders according to the cleared volumes and schedules.

Typically, market participants have non-convex preferences and the allocation (or, welfare maximization) and pricing problems are modeled as mixed-integer linear programs (MILPs). Similar problems can be found in the electricity spot markets organized by Independent System Operators (ISOs) in the U.S. \citep{iso-price-formation}.
Non-convex preferences stem from e.g., startup costs or minimum runtimes of generators which can easily be modeled via binary variables. Non-convex preferences induce non-convex markets and have two main consequences: 
\begin{enumerate}
	\item Walrasian equilibria need not exist in such markets \citep{ahunbay2022pricing}. 
	\item The allocation problem cannot be solved in polynomial time, since MILPs are NP-hard \citep{garey1979computers, burer2012non}. 
\end{enumerate}

Walrasian equilibria are desirable, since they are efficient, envy-free and budget balanced. 
However, these properties hinge on the convexity of the underlying welfare maximization problem. 
In the presence of non-convexities, some of these properties need to be loosened. 
The EU PCR relaxes efficiency and computes an allocation subject to linear and anonymous prices\footnote{Linear and anonymous prices are defined per good and are identical for all market participants.}. In multiple iterations, the algorithm adds constraints to the welfare maximization problem until such prices are found. No bounds on the efficiency loss are known. 
In contrast, U.S. ISO markets enforce the welfare-maximizing solution and compute linear and anonymous prices that violate envy-freeness and budget balance. 
In fact, most ISOs pay out-of-market uplifts (a.k.a. make-whole payments) such that the participants break even, while some even pay lost opportunity costs\footnote{Lost opportunity costs are defined as the difference between each partitipant's profits under the welfare maximizing allocation and the individual profit maxima each participant could obtain given the prices.} such that there is no incentive to deviate from the outcome \citep{ahunbay2022pricing}. 
The U.S. FERC recently argued that ``the use of side-payments can undermine the market’s ability to send actionable price signals'' \citep{ferc-price-formation}. The resulting price signals are considered intransparent and can set flawed incentives as they do not reflect marginal costs anymore.

Another concern with market clearing is computational complexity. MILP solvers have become very effective and even the PCR solution can be computed within 17 minutes today. Yet, there are limits to the scalability of MILPs, and allocation problems will grow larger. 
First, the EU plans to shorten the current market time unit of one hour to 15 minutes \citep{sdac}. 
Second, the zonal pricing rule in the EU day-ahead market simplifies the problem in that most of the network constraints are not considered. If the EU decided to move to nodal pricing, the optimization model would increase substantially\footnote{In a recent nodal simulation organized by ENTSO-E, the consortium of European TSOs, the model included around 25,000 generators, 22,000 lines, 25,000 nodes and 25,000 critical network elements and contingencies. The original mixed-integer linear model was relaxed to a linear optimization models for tractability. See \url{https://www.entsoe.eu/news/2022/06/30/entso-e-publishes-its-report-on-locational-marginal-pricing-study-of-bidding-zone-review-process/}.}. 
Finally, due to the increase in renewable energies, many more market participants are expected, as large fossil or nuclear generators are replaced by smaller wind farms or solar parks. As a result, the problems will grow even larger. At the same time, non-convexities will continue to matter \citep{byers2022}. For example, gas turbines are used to compensate the volatility of wind and solar energy supply, and their activation incurs ramping costs. Furthermore, curtailment costs or transmission and distribution costs that depend on the concentration of renewable energy sources can introduce non-convexities. % This also contributes to the growth of U.S. ISO markets.
The trade-off between efficiency losses of the EU PCR and the uplifts paid in U.S. ISO markets is not well understood\footnote{Note that the EU computes zonal prices whereas the U.S. ISOs determine locational marginal (aka. nodal) prices, which is another major difference between these two large jurisdictions.}. Computational complexity is a concern in both jurisdictions, and for very large markets as in the EU it might become a barrier soon. 

A growing theoretical literature deals with the question, how competitive equilibrium can be approximated in markets with non-convexities \citep{bichler2018matter, bichler2019designing}. In a recent contribution, \citet{Milgrom.2021} proposed the \textit{markup mechanism} to compute approximate Walrasian equilibria in non-convex markets. This mechanism always delivers feasible, near-efficient allocations with no budget deficit, and can be computed efficiently using convex optimization. 
In contrast to earlier work, this approach deviates from a single market price vector to having two such vectors, one for the sellers and one for the buyers. Buyers pay an uplift to cover possible excess supply due to the non-convexities. 
Electricity markets are probably the most relevant applications of this approach. However, the theory draws on existence theorems such as the Shapley-Folkman Lemma \citep{starr1969quasi}, and it is not clear how well such an approach would work in practice and how it would be implemented.

\subsection{Contributions} 

In this paper, we develop an implementation of the markup mechanism for electricity day-ahead markets and experimentally compare its outcome with the incumbent solution, IP pricing.
IP pricing computes the optimal solution of the allocation problem, fixes the integer variables and uses the dual variables of the (exact) demand-supply constraint as prices \citep{oneill2005efficient}. It is known that IP pricing leads to significant make-whole payments \citep{ahunbay2022pricing}, and that relaxing optimality of the solution (allowing for a higher optimality gap) distorts the price signals \citep{byers2022economic}. With larger problem sizes, the allocation problem cannot be solved to optimality, however.
Although different pricing rules have been suggested for electricity markets \citep{liberopoulos2016critical}, in this paper we focus on IP pricing, because it implements the standard approach of marginal-cost pricing, provides accurate congestion signals\footnote{Accurate congestion signals mean that price differences in an electricity network only occur if transmission lines are congested and vice versa.} \citep{ahunbay2022pricing} and it is widely used in U.S. ISO markets\footnote{The EUPHEMIA algorithm \citep{nemo2020euphemia} used in the EU is a complex heuristic that solves a series of mixed-integer programming problems. Details of the implementation are not publicly available.}.

The original markup mechanism allows for excess supply due to the non-convexities in the market and uses two price vectors to avoid a budget deficit. The buyers pay more for each electricity unit to cover the excess supply, and their payments are defined using a non-negative markup. As a result, the market is budget balanced and, under certain conditions (see Section \ref{sec:mw}), can be cleared in polynomial time using convex optimization techniques. 
Moreover, the original markup mechanism computes a relaxation of the allocation problem and assumes that a rounded solution based on the Shapley-Folkman Lemma is available. However, the Shapley-Folkman Lemma is not constructive. Applying the markup mechanism for electricity markets requires constructive techniques to obtain integer-feasible solutions.   
We introduce a polynomial-time rounding method and also one that uses MILP techniques to round the fractional variables (that should be integers in a feasible solution) resulting after solving the relaxation. Given the rounded values, we solve the allocation problem with fixed integer variables to set the continuous variables optimally. This two-stage solution process is different to the original proposal. 
Moreover, the original mechanism includes additional demand units depending on the size of the non-convexities in the agents' preferences. The third example provided in Section \ref{sec:example} shows that these units might render the allocation problem infeasible, and, therefore, we omit them in our implementation. 
Finally, we note that while a solution with excess supply is inadequate for a real-time power market, it might well be feasible for the day-ahead market. The subsequent intraday or real-time markets\footnote{For EU intraday markets see \url{https://www.entsoe.eu/network_codes/cacm/implementation/sidc/}. Information on U.S. real-time markets can be accessed here: \url{https://www.iso-ne.com/markets-operations/markets/da-rt-energy-markets}.} can be used to balance supply and demand and the market operator can provide excess supply from the day-ahead market on these subsequent markets. 
Thus, in our implementation of the markup mechanism for power markets, we also consider solutions with excess supply. 

Based on data sets from the ARPA-E grid optimization competition \citep{arpa-competition} we compute allocations and prices and compare runtimes, the average efficiency loss of the markup mechanism, and the uplifts paid compared to IP pricing. These data sets were prepared to resemble real-world instances and have a fixed number of market participants and a specific network topology. The availability of such public data sets allows for a comprehensive empirical evaluation of the markup mechanism. 
We analyze problems of different size i.e., six power networks between 617 and 18,889 nodes. While the ARPA-E data sets describe allocation problems for a single period, we need to consider multi-period constraints such as minimum uptime requirements of generators in the day-ahead market. Hence, we extend these data sets to multi-period scenarios. This leads to much larger problems that are substantially harder to solve. 
We show that the markup solution is always budget-balanced or has a budget surplus, with appropriate markups. Interestingly, the welfare loss is very small (around $0.01\%$) as is the markup required on the buy-side prices. Another advantage is that the uplifts that need to be paid are substantially lower compared to IP pricing and can be covered by the budget surplus. For instances up to $8,718$ nodes the markup mechanism is significantly faster than the computation of an optimal allocation solution, but both can be solved within an hour. For an instance with 10,480 nodes and one for 12,209 nodes the markup solution can be computed within an hour, but the optimal solution cannot. The 12,209-node instance could not even be solved within 12 hours to optimality. Larger instances with 18,889 nodes available in the ARPA-E data set could neither be solved to optimality nor with the markup mechanism within relevant time frames. 

The paper is structured as follows: we first provide a short review of the related literature on electricity market pricing. Then we introduce a model of a non-convex electricity market and a short description of widely used payment rules in Section 3. In Section 4, we summarize the markup mechanism and the necessary adaptations in the context of electricity day-ahead markets. Section 5 introduces the data sets and summarizes the experimental results, before we conclude in Section 6.

\section{Related Literature} \label{sec:literature}

In this work, we focus on wholesale electricity markets and on day-ahead markets in particular. In these markets, participants submit buy and sell bids using a specific bid language, which then translates into a central welfare maximization problem. As indicated above, bid languages typically involve some form of non-convexities e.g., by considering start-up costs, ramping constraints, or block orders of market participants \citep{herrero2020evolving}. These non-convexities are modelled as MILPs \citep{Hillier.2001}, which are NP-hard problems \citep{garey1979computers}. Previous contributions have identified the complexity of the allocation problem in electricity markets and various alternative models have been proposed for their solution, including Lagrangian relaxations (which were used before MILPs in U.S. power markets), or decomposition techniques \citep{Galiana.2002,Padhy.2004,Zheng.2016}. 

Aside from the complexity to solve the allocation problem, computing prices in the presence of non-convexities is a fundamental challenge. The welfare theorems show that Walrasian equilibrium prices exist in convex markets \citep{arrow1954existence}. Walrasian equilibria are efficient, envy-free, and budget-balanced. However, they do not exist in non-convex markets in general \citep{Kelso82, bikhchandani1997competitive, baldwin2019understanding}. 
Electricity markets are a prime example for non-convex markets where we cannot expect Walrasian equilibrium to exist. Different pricing rules were suggested to deal with these non-convexities \citep{liberopoulos2016critical}. We focus on Integer Programming (IP) and Convex Hull (CH) pricing as the two pricing rules that have received most attention. IP pricing is widely used in U.S. ISO markets. 
CH pricing \citep{hogan2003minimum, gribik2007market} aims to approximate Walrasian equilibria by finding prices that minimize lost opportunity costs. CH pricing is based on replacing non-convex valuation functions by their convex hulls and obtaining prices from the dual of the resulting problem. It is computationally hard \citep{Schiro.2016}. Extended Locational Marginal pricing (ELMP) derives prices from the linear programming relaxation of the allocation problem. Under certain assumptions on the bidding formats it corresponds to Convex Hull pricing \citep{Hua.2017}. ELMP pricing is currently used for fast-start pricing in MISO and ISO-NE \citep{MISO.2019,iso-ne-fast-start-pricing}. 
Most ISO power markets in the U.S. are based on IP pricing \citep{oneill2005efficient}. Here, prices are obtained from the dual of a convex problem where the integer variables are fixed to the values they attain under the optimal allocation. 
In both pricing rules, market participants bear lost opportunity costs \citep{ahunbay2022pricing}. 

The question how linear and anonymous prices can be computed in two-sided non-convex markets has received attention beyond electricity markets alone. \citet{bichler2018matter} suggest to compute constrained efficient allocations subject to the availability of linear and anonymous prices. However, their mechanism requires solving a MILP. 
\citet{Milgrom.2021} provide the markup mechanism for allocation and pricing that only require convex optimization. 
In what follows, we introduce this algorithmic framework in detail and discuss challenges that arise in electricity markets.

\section{Pricing on non-convex markets}\label{sec:preliminaries}

In this section, we present our model for coupled markets and the specific model of electricity market clearing we consider throughout the paper. We next discuss the theory of clearing convex and non-convex markets, and the notion of (approximate) equilibria. 

In what follows, we denote by $\extR = \mathbb{R} \cup \{-\infty,\infty\}$ the extended real line, where addition and multiplication are commutative binary operations defined as usual over the reals, and are extended such that for any $x \in \mathbb{R}$, $x + \infty = \infty$, $x - \infty = -\infty$. For any function $f : \mathbb{R}^d \rightarrow \extR$, its domain $\textnormal{dom}(f) = \{x \in \mathbb{R}^d | f(x) \in \mathbb{R}\}$ is the subset of $\mathbb{R}^d$ on which $f$ does not equal $\pm \infty$. A function $f : \mathbb{R}^d \rightarrow \extR$ is said to be (1) convex if for any $\lambda \in [0,1]$ and any $x, y \in \mathbb{R}^d$, $f(\lambda x + (1-\lambda)y) \leq \lambda f(x) + (1-\lambda) f(y)$, (2) lower semi-continuous (closed) if for any $x \in \mathbb{R}^d, \lim \inf_{y \rightarrow x} f(y) \geq f(x)$, and proper if both $f(x) \neq -\infty$ for any $x \in \mathbb{R}^d$ and $f \neq \infty$ identically. When a function $f$ is not closed or convex, we may consider the best such ``approximation'' of $f$. Specifically, we denote by $\ch(f)$ the convex closure of $f$, the pointwise maximum closed and convex function which underestimates it. Analogously, a function $f$ is concave if $-f$ is convex, upper semi-continuous if $-f$ is closed, and the concave closure $\conch(f)$ of $f$ is simply defined as $-\ch(-f)$. 

We call a subset $S$ of $\mathbb{R}^d$ convex or closed if its characteristic function $\chi_S$, defined for any $x \in \mathbb{R}^d$ as $\chi_S(x) = 0$ when $x \in S$ and $\chi_S(x) = \infty$ otherwise, is convex or closed. Furthermore, for any subset $S \subseteq \mathbb{R}^d$, we denote by $\conv S = \{ \sum_{k = 1}^{d+1} \lambda_k x_k | \forall \ k, \lambda_k \geq 0 \textnormal{ and } x_k \in S, \sum_{k=1}^{d+1} \lambda_k = 1\}$ the convex hull of $S$. Finally, given an indexed set $(S_i)_{i \in I}$ of subsets of $\mathbb{R}^M$, we denote by $+_{i \in I} S_i = \{ \sum_{i \in I} s_i | \forall \ i \in I, s_i \in S_i\}$ the Minkowski sum of $(S_i)_{i \in I}$.

\subsection{Market model} \label{subsec:notation}
We consider a two-sided coupled electricity market given by (1) a network containing nodes connected via transmission lines, and (2) agents eliciting preferences and constraints. 

In full generality, the model has a set of buyers $B$, a set of sellers $S$, and a set of transmission operators $R$, and we denote by $L = B \cup S \cup R$ the set of agents. There are $M$ goods, spatially and temporally differentiated i.e., two copies of same good traded at a different place and time are encoded as distinct goods. Each buyer $b$ has a valuation function $v_b : \mathbb{R}^M_+ \rightarrow \extR$, and each seller $s$ has a cost function $c_s : \mathbb{R}_+^M \rightarrow \extR$. Transmission operators, meanwhile, are assumed to enact their trades at zero cost; each transmission operator $r$ has a cost function $d_r : \mathbb{R}^M \rightarrow \{0,\infty\}$. We denote by $\ell \in L$ a generic agent, and $c_\ell$ the associated cost function of $\ell$, where we set $c_\ell = -v_\ell$ if $\ell \in B$. An economy $\mathcal{E}$ is then a tuple $\left(B,S,R,(v_b)_{b \in B}, (c_s)_{s \in S}, (d_r)_{r \in R}\right)$. 

We often denote by $x_b$ the vector of goods consumed by buyer $b$, $y_s$ the vector of goods supplied by seller $s$, and $f_r$ the vector of exchanges enacted by transmission operator $r$. For a generic agent $\ell$, we denote their bundle as $z_\ell$, where we set $z_\ell = -x_\ell$ if $\ell \in B$. An allocation is then a tuple $(z_\ell)_{\ell \in L} = \left( (-x_b)_{b \in B}, (y_s)_{s \in S}, (f_r)_{r \in R} \right)$, which we may also denote simply as $z$. 

Given an allocation $(z_\ell)_{\ell \in L}$, the vector of excess supply is defined as 
\begin{equation}
    \sigma(z) = \sum_{s \in S} y_s - \sum_{b \in B} x_b + \sum_{r \in R} f_r.
\end{equation} 
The notion of feasibility of an allocation depends on whether we demand weak or strict supply-demand equivalence. Specifically, we say that weak (strict) supply-demand equivalence holds for an allocation $z$ if $\sigma(z) \geq (=) \ 0$, and an allocation is weakly / strictly feasible if both $z_\ell \in \textnormal{dom}(c_\ell)$ for every agent $\ell$ and the associated notion of supply-demand equivalence holds. We denote by $\hat\Omega \ (\Omega)$ the set of weakly (strictly) feasible allocations.

In our consideration of electricity markets, we employ a simplified multi-period \emph{direct current optimal flow} (DCOPF) \citep{Hua.2017} model with lossless transmission. The transmission grid is represented as a graph with nodes $V$, a reference node $R^* \in V$ and edges $E$, and we denote by $N(v) = \{w \in V | (v,w) \in E\}$ the set of neighbouring nodes to $v$. The set of goods for the market are then separated into two categories, electric power at node $v \in V$ at time $t$, and line flow on edge $(v,w) \in E$ at time $t$. For each edge $e=(v, w) \in E$, the minimum/maximum flow capacity ($\underline{F}_{vw}/ \overline{F}_{vw}$) and the susceptance $B_{vw}$ are specified. The feasible trades are then specified by line flows $f_{vwt}$ on the edge $(v,w)$ at time $t$ and phase angles $\alpha_{vt}$ at node $v$ at time $t$, satisfying inequalities
\begin{align}
    f_{vwt} - B_{vw}(\alpha_{vt} - \alpha_{wt}) & = 0, \label{cons:flow-dem} \\
    \underline{F}_{vw} \leq f_{vwt} & \leq \overline{F}_{vw} \ \forall \ v \in V, w \in N(v), t \in T.
\end{align}
In particular, the network characteristics are assumed to be static over each time period $t \in T$. To represent feasible trades enacted by transmission operators, we consider a single transmission operator whose feasible trades are characterised by the constraints on $f_{vwt}$ and $\alpha_{vt}$. In particular, at time $t$, the transmission operator supplies $\sum_{w \in N(v)} f_{wvt}$ units of power to node $v$.

Furthermore, each node potentially contains one or more agents. The agents are given by two disjoint sets, $S$ (sellers) and $B$ (buyers). Each seller $s \in S$ is mapped to a specific node $\nu(s)$ in the network and is associated to a set of bids $\beta_s^t$ for each time period $t \in T$. Every bid $l \in \beta_s^t$ has the form $(q_{stl}, c_{stl})$, where $q_{stl}$ is the maximum quantity (in MWh) that can be produced and $c_{stl}$ denotes the variable production costs. Each seller may be subject to several constraints, such as minimum output ($\underline{P}_{st}$), no-load costs ($h_s$) or minimum uptime ($\underline{R}_{s}$). Non-convexities that arise from such constraints are then often encoded via additional binary variables. The term $c_s(y_s)$ denotes the total cost inquired by $s \in S$ to produce the amount $y_s$.

Buyers $b \in B$ are similarly mapped to a specific node $\nu(b)$ in the network and have a set of block bids $\beta_b^t$ for every time period $t \in T$. A block $l \coloneqq (q_{btl}, v_{btl})$ consists of a maximum quantity $q_{btl}$ and a valuation $v_{btl}$. 
Moreover, for each period $t$, a buyer has a minimum demand ($\underline{P}_{bt}$) i.e., a price-inelastic demand that must be satisfied in each feasible allocation.
The total value of buyer $b\in B$ to consume quantity $x_b$ is denoted as $v_b(x_b)$.

We usually require strict supply-demand equivalence for feasibility. In particular, we interpret (\ref{cons:flow-dem}) as a strict feasibility constraint for flow goods, and also ask for equality of supply and demand of power at each node,
\begin{equation}
    \sum_{\nu(s) = v} y_{st} - \sum_{\nu(b) = v} x_{bt} - \sum_{w \in N(v)} f_{wvt} = 0 \ \forall v \in V, t \in T.
\end{equation}
A detailed mathematical model (\hyperref[opt:dcopf]{DCOPF-MILP}) and a list of symbols are provided in Appendix \ref{app:dcopf}. 

\subsection{Pricing, utilities, and equilibria}

A price vector $p \in \mathbb{R}^M$ denotes the payment an agent makes per unit of good they purchase or supply. We assume that agents have quasilinear utilities, and given an allocation, agents derive a payoff equal to the transfer they receive minus their costs. Formally, for each buyer $b$, seller $s$ and transmission operator $r$, the utilities are defined as 
\begin{align}
    u_b(x_b,p) & = v_b(x_b) - p^T x_b, \\
    u_s(y_s,p) & = p^T y_s - c_s(y_s), \\
    u_r(f_r,p) & = p^T f_r - d_r(f_r).
\end{align}

Given a price vector, an agent may consider maximizing their payoff amongst a subset of allocations feasible for them. Towards this end, we denote an agent $\ell$'s indirect utility relative to a subset $X \subseteq \mathbb{R}^M$ as $\hat{u}_\ell(X,p) = \max_{z_\ell \in X} u_\ell(z_\ell,p)$. Agent $\ell$'s demand correspondence relative to $X$, $D_\ell(X,p) : 2^{\mathbb{R}^M} \times \mathbb{R}^M \rightrightarrows \mathbb{R}^M$, is the set of utility maximizing bundles for agent $\ell$ within $X$, $D_\ell(X,p) = \arg \max_{z_\ell \in X} u_\ell(z_\ell,p)$. Whenever $X = \mathbb{R}^M$, we simplify notation by dropping reference to $X$.

Given an economy $\mathcal{E}$, the welfare associated with an allocation $z$ is the quantity 
\begin{equation}
    \label{eq:welfare}
    W(z) = \sum_{b \in B} v_b(x_b) - \sum_{s \in S} c_s(y_s) - \sum_{r \in R} d_r(f_r).
\end{equation}
Then the welfare maximization problem with respect to weak (strict) feasibility is defined as
\begin{equation}\label{opt:welfare} 
    \max_{x,y,f} W(z) \textnormal{ subject to }  z \in \hat{\Omega} \ (\Omega).
\end{equation}
A solution of (\ref{opt:welfare}) is called an efficient allocation, and is often denoted by $z^*$ or by $(z^*_\ell)_{\ell \in L} = \left( (x^*_b)_{b \in B}, (y^*_s)_{s \in S}, (f^*_r)_{r \in R} \right)$. Then given any allocation $z$, we consider the relative percentage welfare loss (RWL) as a measure of optimality. RWL is equal to $\%100 \cdot (W(z^*) - W(z)) / W(z^*)$.

An allocation and price pair $(z,p)$ thus allows us to compute the total gains from trade, as well as participants' payoffs; in other words, it specifies an outcome of the market. Some desirable properties of such an outcome are:
\begin{enumerate}
    \item \textbf{Efficiency:} $W(z)$ equals the value of (\ref{opt:welfare}).
    \item \textbf{Supply-Demand Balance:} $z$ is feasible for (\ref{opt:welfare}).
    \item \textbf{Envy-Freeness:} $z_\ell \in D_\ell(p)$ for every agent $\ell$.
    \item \textbf{Individual Rationality:} $u_\ell(z_\ell, p) \geq 0$ for every agent $\ell$.
    \item \textbf{Budget Balance:} The auctioneer provides no subsidy or makes no profit i.e., 
    $$ p^T \left( \sum_{b \in B} x_b - \sum_{s \in S} y_s - \sum_{r \in R} f_r \right) = 0. $$
\end{enumerate}
Such an allocation-price pair $(z,p)$ is called a Walrasian (or, competitive) equilibrium. When an economy $\mathcal{E}$ is convex i.e., when all valuations are concave and closed and when all costs are convex and closed, the celebrated result of \citet{arrow1954existence} characterizes when a Walrasian equilibrium exists.

\begin{proposition}
    Let $\mathcal{E}$ be a convex economy. Then an allocation $z$ is an optimal solution of (\ref{opt:welfare}) if and only if there exists prices $p$ such that $(z,p)$ is a Walrasian equilibrium.
\end{proposition}

\citet{ahunbay2022pricing} provide a version of this theorem for a set of convex markets that are connected via a transmission network. They assume quasi-linear utility functions, as in this paper, and the proof is via the Fenchel-Young inequality from convex analysis. 

\subsection{Allocation and pricing in non-convex (electricity) markets}\label{sec:preliminaries_allocation_pricing}

Whereas convexity is a standard assumption in classical economics, in many real world settings, such as spectrum, transportation, or environmental access markets, it is violated. In the concrete setting of electricity markets, constraints such as minimum generation levels and minimum uptimes provide sources of non-convexities. In a non-convex market, a Walrasian equilibrium need not exist. This raises the question of how to clear such markets, or namely, how to find an allocation and price pair which provides an outcome satisfactory to the market designer.

In the context of electricity markets, market operators in the U.S. typically compromise on envy-freeness. Here, an optimal solution to the welfare maximization problem (\ref{opt:welfare}) is computed, and linear and anonymous prices $p$ for this optimal allocation are computed afterwards. Finding the \emph{``correct''} pricing rule for these markets is an open problem, and many pricing rules have been proposed over the years. For a detailed survey we refer the reader to \citep{liberopoulos2016critical}. We note, however, that the more prominent pricing rules often minimize a measure of deviation from Walrasian equilibria. Namely, given an allocation $z$ and a price vector $p$,
\begin{enumerate}
    \item The \textbf{(global) lost opportunity cost} of an agent $\ell$ measures the difference between the agent's indirect utility and their realised utility, $\textnormal{GLOC}_\ell(z_\ell,p) = \hat{u}_\ell(p) - u_\ell(z_\ell,p)$.
    \item The \textbf{make whole payment} of an agent $\ell$ equals the amount by which their \emph{individual rationality} constraint is violated given their allocation and the prices. Formally, $\textnormal{MWP}_\ell(z_\ell,p) = \hat{u}_\ell(\{z_\ell,0\},p) - u_\ell(z_\ell,p)$.
\end{enumerate}
The major pricing rules used in U.S. electricity markets are based on CH pricing and IP pricing. These pricing rules respectively correspond to prices which minimize the sum of agents' global lost opportunity costs \citep{hogan2003minimum} and local lost opportunity costs (a subset of GLOCs, see \citep{ahunbay2022pricing}) \citep{oneill2005efficient}, for an optimal allocation $z$.

Market operators typically enforce stability via penalties and they do not pay GLOCs. Penalites avoid deviations from the welfare maximization allocation. Participants are usually paid MWPs such that the individual rationality of each market participant is guaranteed. The magnitude of these payments have been a source of concern, and to rectify this, \citet{Bichler.2021} proposed a pricing rule which minimizes the sum of participants' make-whole payments. Another design goal of a pricing scheme is having adequate \textit{congestion signals}. This means that the prices at two neighboring nodes $v$ and $w$ should be equal if the network is not congested i.e,. $f_r$ is in the interior of $\textnormal{dom}(d_r)$ for each transmission operator $r$. %Congestion signals are strongly connected to LLOCs, and whenever transmission operators incur LLOCs they can incur a product revenue shortfall \citep{Schiro.2016,ahunbay2022pricing}. Thus, to obtain a more balanced tradeoff between LLOCs and MWPs, \citet{ahunbay2022pricing} proposed a pricing rule which minimizes the sum $\sum_{\ell \in L} \max\{\textnormal{MWP}_\ell(z_\ell,p), \textnormal{LLOC}_\ell(z_\ell,p)\}$, which is dubbed as the \emph{join} of IP and min-MWP pricing. %\JK{Guess in section 4 or 5 we could provide a bit more details how these pricing rules formulate?} \MA{Do this in the appendix, then we check whether we have enough space.}\MB{Agree}

In the European electricity market, the bidding language allows for the expression of convex valuations in the form of hourly bids, and non-convex valuations via so called block bids and (scalable) complex orders. The market clearing, in turn, trades off efficiency to obtain prices that satisfy envy-freeness to a greater extent than that of U.S. electricity markets. Specifically, market clearing necessitates that linear prices are imposed at each zone for an allocation, such that (1) hourly bids are accepted if and only if they are in- or at-the-money, and (2) bids representing non-convex valuations are cleared only if they are not out-of-the-money. Abstracting away from ideosyncracies of the European market clearing e.g. the specific requirements for clearing the Italian market or the inclusion of long term allocations, the European market clearing problem may be written as
\begin{align}
    \max_{z,p} \ & W(z) \textnormal{ subject to } & z & \in \Omega \\
 && \textnormal{GLOC}_\ell(z_\ell,p) & = 0 \ \forall \ \ell \in L \textnormal{ such that } c_\ell \textnormal{ is convex} \nonumber \\
 && \textnormal{MWP}_\ell(z_\ell,p) & = 0 \ \forall \ \ell \in L \textnormal{ such that } c_\ell \textnormal{ is non-convex.} \nonumber
\end{align} 

An exact implementation of this optimization problem leads to non-linearities. The PCR EUPHEMIA \cite{nemo2020euphemia} describes a series of mixed-integer linear programs aiming to solve the optimization problem while considering specifics of certain European price zones. Note, however, that in contrast to U.S. ISO markets, the EU Single Day-Ahead Market (SDAC) compromises on welfare maximization. The markup mechanism by \citet{Milgrom.2021} that we describe next, also relaxes the welfare maximization goal, but provides bounds on the welfare loss and it can be computed via a series of convex optimization problems as long as the rounding can be done efficiently.

\section{The markup mechanism}

While Walrasian equilibria might not exist in the presence of non-convexities, economic intuition suggests that in large markets, such non-convexities are \emph{smoothed} out and an allocation and price pair which forms a \emph{``near''} Walrasian equilibrium exists. It was in \citet{Starr.1969} where this notion was made explicit, and recently, \citet{Milgrom.2021} adapted the formalism therein to provide a mechanism which outputs a ``two-price'' equilibrium for an economy under weak supply-demand equivalence. 

\subsection{Algorithmic framework}\label{sec:mw}

The \emph{quality} of the notions of approximate equilibria \citep{Starr.1969}, in terms of its approximate feasibility or approximate envy-freeness, depends on an appropriately defined measure of non-convexity for an economy. To wit, for a set $S \subseteq \mathbb{R}^d$, the \textbf{radius} of $S$ is given by $\rad S = \inf_{x \in \mathbb{R}^d} \sup_{y \in S} \| x - y\|$, where $\| \cdot \|$ is the standard Euclidean norm on $\mathbb{R}^d$. For our measures of non-convexity, we denote by $r(S) = \sup_{x \in \conv S} \inf_{T \subseteq S | x \in \conv T} rad(T)$ the \textbf{inner radius} of $S$, and by $\rho(S) = \sup_{x \in \conv S} \inf_{y \in S} \| x - y \|$ the \textbf{inner distance} of $S$. Then for an economy $\mathcal{E}$, we define our measures of non-convexity as
\begin{align}
    r_\mathcal{E} & = \max_{\ell \in L} \sup_{p \in \mathbb{R}^M} r\left(D_\ell(p)\right), \\
    \rho_\mathcal{E} & = \max_{\ell \in L} \sup_{p \in \mathbb{R}^M} \rho\left(D_\ell(p)\right).
\end{align}
Of course, buyers' valuations are concave and sellers' and transmission operators' costs are convex and closed if and only if the demand correspondences of agents are closed and convex valued for any price vector $p$. Thus $r_\mathcal{E}, \rho_\mathcal{E}$ are non-zero valued if and only if there is an agent with non-convex preferences. In this case, the Shapley-Folkman Lemma and its extensions due to \citet{Starr.1969} and \citet{Heller.1972}\footnote{The lemma itself and its corollaries are often collectively referred to as the Shapley-Folkman Lemma.} provides guarantees on approximate equilibria.

\begin{lemma}[Shapley-Folkman Lemma]
    For $M \in \mathbb{N}$, suppose that $S_i \subseteq \mathbb{R}^d$ for each $i = 1, 2, \ldots, M$, and consider $S = +_{i = 1}^M S_i$. Then, denoting $K = \min\{d,M\}$, for any $x \in \conv S$, $x = \sum_{i=1}^{M} x_i$ such that $x_i \in \conv S_i$ for every $i$. Furthermore, except for at most $K$ many $i$, $x_i \in S_i$.
\end{lemma}

\begin{corollary}[\citep{Starr.1969}, Appendix 2]
    In the setting of the Shapley-Folkman Lemma, there exists $y \in S$ such that $\| x - y \| \leq \sqrt{K} \cdot \max_{1 \leq i \leq M} r_{S_i}$.
\end{corollary}

\begin{corollary}[\citep{Heller.1972}]
    In the setting of the Shapley-Folkman Lemma, there exists $y' \in S$ such that $\| x - y' \| \leq K \cdot \max_{1 \leq i \leq M} \rho_{S_i}$.
\end{corollary}

To obtain guarantees on approximate equilibria, we then consider the convexified economy $\conv \mathcal{E} = \left(B,S,R,(\conch \ v_b)_{b \in B}, (\ch \  c_s)_{s \in S}, (\ch \ d_r)_{r \in R}\right)$. If $(z,p)$ is a Walrasian equilibrium for this economy, then by applying the Shapley-Folkman Lemma to the Minkowski sum of agents' demand sets, $z$ may be chosen such that agents are assigned a bundle in their demand set -- except at most $\min\{M,L\}$ many who are instead assigned bundles infeasible or undemanded for them. Such a pair $(z,p)$ is called a \textit{pseudoequilibrium}. Furthermore, by the corollaries of \citet{Starr.1969} and \citet{Heller.1972}, there is an allocation $z'$ of bounded Euclidean distance to $z$ such that all agents are utility maximizing but potentially $z'$ violates supply-demand equivalence; such a pair is instead denoted a \textit{quasiequilibrium} or approximate equilibrium.

\citet{Milgrom.2021} leverage the existence of quasiequilibrium in the construction of their \emph{simple markup mechanism} (for a market without transmission). The mechanism first checks whether a Walrasian equilibrium exists. If not, to compute an allocation, the mechanism considers the \emph{scaled} welfare for an allocation $z$ and scale factor $\alpha$,
\begin{equation}
    \widehat{W}(z,\alpha) = \sum_{b \in B} \frac{v_b(x_b)}{1+\alpha} - \sum_{s \in S} c_s(y_s).
\end{equation}
For any $\alpha \geq 0$, the mechanism considers maximizing the scaled welfare for the convexified economy, subject to a weak feasibility constraint modified to leave an excess $R = \min\{ r_\mathcal{E} \sqrt{M}, \rho_\mathcal{E} M\}$ of each good. In particular, the feasibility constraint $\Omega(R)$ is now 
\begin{equation}\label{cons:weak-SD-R}
    \sum_{s \in S} y_s - \sum_{b \in B} x_b \geq R \Sigma, \tag{$\Omega(R)$}
\end{equation}
where $\Sigma$ is the $M$-dimensional all-ones vector. For a Walrasian equilibrium $(z'_\alpha,p^*_\alpha)$ of this convexified economy, by the Shapley-Folkman Lemma there is an allocation $z^*_\alpha$ which satisfies weak supply-demand equivalence. Moreover, for any buyer $b$, $x_b \in D_b(p^*_\alpha(1+\alpha))$, and for any seller $s$, $y_s \in D_s(p^*_\alpha)$. Therefore, we may charge buyers price $p^*_\alpha(1+\alpha)$ and the sellers price $p^*_\alpha$ to ensure that all market participants obtain an allocation in their demand set given the prices they face. Whereas a good whose supply-demand constraint does not bind for the allocation $z$ might have a positive price, the markup $\alpha$ offers the possibility that excess payments from buyers may cover the budget deficit. The mechanism thus considers the set of markups $\alpha$ which do satisfy budget balance, 
$$A(\mathcal{E}) = \left\{ \alpha \geq 0 \ \Bigg| \ p^{*T}_\alpha \left(\sum_{b \in B} (1+\alpha) x^*_{b\alpha} - \sum_{s \in S} y^*_{s\alpha} \right) \geq 0 \right\},$$
and picks the minimum markup $\alpha^* = \min A(\mathcal{E})$ which achieves budget balance.

Finally, \citet{Milgrom.2021} show that for a sufficiently large economy, the set $A(\mathcal{E})$ is closed and non-empty, and a minimum markup exists. Moreover, relative efficiency loss and the markup $\alpha^*$ both decline linearly in the size of the market. This implies the existence a two-price mechanism which requires solving only convex optimization problems to approximately clear a market in the presence of non-convexities. Specifically, modulo considerations regarding the choice of $\alpha$, a budget balanced market clearing is achieved by solving a \emph{single} convex optimization problem. 

The electricity markets for which we implement the markup mechanism are distinct in several ways from the large economies considered by \citet{Milgrom.2021}. Some key differences are:
\begin{enumerate}
    \item The electricity markets we consider possess not only sellers and buyers, but also transmission operators, who may exchange one good for another.
    \item The existence and efficiency results are proven for sufficiently large markets. That is, for a sequence of economies it is assumed that the number of participants grows, but not the number of items. Therefore, the relative number of items to market participants is vanishing.  While the scale of a practical electricity market is large in the number of participants, electric power delivered to each node at each time period is encoded as a distinct good. Therefore, the number of goods can also be significant compared to the number of buyers and sellers.
		\item Constructive rounding procedures that lead to integer feasible solutions that satisfy the corollaries by \citet{Heller.1972} and \citet{Starr.1969} are not readily available.
		\item Electricity markets in practice demand a strict supply-demand equivalence.
\end{enumerate}
The first difference raises the question of what prices the transmission operators should face, and whether the efficiency guarantees of \citet{Milgrom.2021} persist. This question turns out to be merely technical, and we remark that an identical argument to theirs shows that transmission operators may be cleared at either the buyers' or the sellers' prices. 

The second and third points, on the other hand, have implications with respect to clearing and rounding. 
\citet{Milgrom.2021} show that there exists an $R$ and there exists a rounding such that all participants get an element of their demand or supply set. % see page 20 of their paper
However, there are many possible roundings and just finding a feasible rounding is NP-hard in general.
For the standard DCOPF used in our work, there exists a polynomial-time threshold-based rounding as long as no complex constraints (e.g., maximum uptime) are added. However, this rounding does not satisfy the corollaries by \citet{Heller.1972} and \citet{Starr.1969}. The lack of a readily available $R$ and a polynomial-time rounding scheme that satisfies these properties has consequences for the properties that can be achieved with the MW framework. 
Given the multitude of items, we stick to an auctioneer demand of $R=0$ and a threshold-based rounding. An auctioneer demand of $R=0$ leads to lower total demand and thus lower dual prices. 
With the threshold-based rounding, some of the buyers might not get an element of their demand set and even make a loss at the prices. However, with an appropriate $\alpha$ we can always find a solution such that budget surplus is high enough to cover the resulting MWPs. 

The final point is based on the physics that govern the market. At least at gate closure supply and demand need to be balanced. However, as stated, excess supply in the day-ahead market may be cleared in the intraday or in a real-time market. We therefore address the final point by allowing for limited excess supply in our relaxed clearing. 
We elaborate on all these issues in Section \ref{sec:example}, while Section \ref{sec:implementation} discusses the specifics of our implementation.

\subsection{Illustrative examples}\label{sec:example}

In what follows, we illustrate the original markup mechanism in the context of an electricity market model. 

\begin{example} \label{ex:vanillaMW}
	Consider an economy with two sellers $s_1, s_2$ and one buyer $b$ at a single node. The minimum production levels are 10 MW for $s_1$ and 8 MW for $s_2$. Seller $s_1$ can produce 10 MW at a marginal cost of \euro{5}/MW and 5 additional MW at a marginal cost of \euro{7}/MW. Similarly, $s_2$ can produce 8 MW at a marginal cost of \euro{4}/MW and 2 additional MW at a marginal cost of \euro{100}/MW. The buyer $b$ has a price-inelastic demand of 8 MW and an elastic demand of 2 MW with associated valuation \euro{10}/MW. 
\end{example}

Note that the buyer has a concave valuation function, while the sellers have non-convex cost functions. We obtain the convex hulls of the cost functions by converting the binary commitment decision variables $u_1, u_2$ to real variables.
Second, we compute the largest inner radius and inner distance (i.e., 10MW/2) of the demand correspondence sets i.e., $r_\mathcal{E} = \rho_\mathcal{E} = 5$. The auctioneer demand is $R=\min\{r_\mathcal{E}, \rho_\mathcal{E}\} = 5$. Then, given some $\alpha \geq 0$, we formulate the following convexified scaled welfare maximization problem which describes a perturbed economy: 
\begin{align*}
	\max_{x, y, u} \ \frac{10}{1 + \alpha}x_b - (5y_{11} + 7y_{12}) - (4y_{21} + 100y_{22}) \\ 
	\text{s.t.} \ \  (y_{11} + y_{12}) + (y_{21} + y_{22}) - (x_b + 8) \geq R \\
	(y_{11} + y_{12}) - 15 u_1 \leq 0 \\
	(y_{11} + y_{12}) - 10 u_1 \geq 0 \\
	(y_{21} + y_{22}) - 10 u_2 \leq 0  \\
	(y_{21} + y_{22}) - 8 u_2 \geq 0 \\
	y_{11} \leq 10u_1, \; y_{12} \leq 5u_1, \; y_{21} \leq 8u_2, \; y_{22} \leq 2u_2 \\
	x_b \leq 2 \\
	u_1, u_2 \geq 0.
\end{align*}
Solving this problem for a given $\alpha$ yields an allocation  $z^{\alpha} = (x^{\alpha},y^{\alpha})$. The seller prices $p^{\alpha}$ are obtained as Lagrangian multiplier of the weak supply-demand balance constraint and buyer prices as $(1+\alpha)p^{\alpha}$. Identifying the optimal (minimal) markup $\alpha^*$ requires search. Thus, we search for the smallest value in $\{0, 0.01, 0.1, 0.2, 1, 1.5\}$ to find some markup $\alpha$ that guarantees no budget deficit. The resulting pair $(z^{\alpha}, p^{\alpha})$ is a pseudoequilibrium of the perturbed economy. 

With $\alpha = 1$ we obtain $x_b^{\alpha} = 0$, $y_{11}^{\alpha} = 5, y_{12}^{\alpha} = 0$, $y_{21}^{\alpha} = 8, y_{22}^{\alpha} = 0$ with $u_1^{\alpha} = 0.5$ and $u_2^{\alpha} = 1$. The buyer faces a price of $(1 + \alpha)p^{\alpha} = 10$ and the sellers are paid at a price $p^{\alpha}=5$. We note that the fractional variable $u_1^{\alpha}$ leads to the infeasibility of the original welfare maximization problem (WMP). For this simplified model, there exists a rounding of the allocation $z^{\alpha}$ based on the Shapley-Folkman Lemma to obtain an approximate (or a quasi-equilibrium) allocation $z'$. This approximate allocation has at most $2R$ units of excess supply
and satisfies feasibility, weak budget-balance, and individual rationality. 
For this, we simply round down the fractional variable $u_1^{\alpha}$ to 0. % and round up the variable $u_2^{\alpha}$ to 1.
The final allocation $z'$ has thus $x'_b = 0$, $y'_{11} = 0$, and $y'_{21} = 8$. In particular, there is no excess supply.
With the two prices $p^{\alpha} = 5$ for sellers and $(1 + \alpha)p^{\alpha} = 10$ for buyers, envy-freeness is achieved for every market participant. Allocation and prices could be computed in polynomial time and zero GLOCs or MWPs occur for market participants.
The optimal allocation is $x_b^* = 2$ and $y_{11}^* = 10$, with a welfare of \euro{-30}. The rounded allocation admits a welfare of \euro{-32} and thus implies an efficiency loss of 6.7\%.
We note, however, that finding a rounded allocation that adheres to the premises of the Shapley-Folkman Lemma is, in general, not trivial, as the following example shows.

\begin{example}
	Consider the setup in Example \ref{ex:vanillaMW}, with the only difference that buyer $b$ now has an inelastic demand of 10 MW.
\end{example}

With $\alpha = 1$, the pseudoequilibrium allocation looks as follows: $x^{\alpha}_b = 0$, $y^{\alpha}_{11} = 7$, $y^{\alpha}_{12}=0$, $y^{\alpha}_{21} = 8$, $y^{\alpha}_{22} = 0$ with $u^{\alpha}_1 = 0.7$ and $u^{\alpha}_2 = 1.0$. 
Then for any threshold $> 0.7$, we round down $u_1$ to 0 and round up $u_2$ to $1$. The rounded allocation $z'$ has $x'_b = 0$, $y'_{11} = 0$, $y'_{21} = 8$, and $y'_{22} = 2$. Thus the Euclidean norm between $\lVert z' - z^{\alpha} \rVert = 7.28 > 5 = R$. Therefore, for this example, the specified rounding is not a candidate rounding according to the Shapley-Folkman Lemma. The excess demand of $7$ after rounding can only partially be offset by the auctioneer demand and the second seller has to compensate for the excess demand as well. Note that through the rounding  we  lose  $7$  supply  units of buyer $1$, creating an overdemand of $7$. This leads to high costs for seller $2$ and this seller, in fact, makes a loss at the specified prices. Finding a rounding that complies with the Shapley-Folkman Lemma is NP-hard in general. In our experiments we use a simple threshold-based rounding (see Section \ref{sec:implementation}). If the chosen threshold does not lead to a feasible solution, we use another threshold. This approach turned out to work well in our experiments. There might be cases with complex buy-side bids where threshold-based rounding does not lead to feasible solutions. We also experimented with MILP-based rounding where all but the fractional variables in the solution to the LP relaxation are determined via mixed integer programming. This approach is more costly, however.

Another problem arises from the many goods traded in electricity markets. 
One of the key ideas of the markup mechanism is to "reserve" a fictitious auctioneer demand $R$ in the convexified market that can be used to offset deviations during rounding. As a result, even after rounding there would be sufficient supply to cover the demand. 
Importantly, this auctioneer demand is added to all of the items. 
In nodal electricity markets, an item equals electricity at a certain point in time at a certain node in the transmission network. A multi-period market clearing of large electricity networks thus implies a large number of items\footnote{For example, PJM has approximately 10,000 pricing nodes, resulting in 240,000 items for the day-ahead market.}. As a result, auctioneer demand $R$ might be unreasonably high, as the following example illustrates: 

\begin{example}\label{ex:infeasible_CSWMP}
	Consider a transmission network represented as a complete graph with three nodes $V_1, V_2, V_3$. In every node $V_i$, there exists one buyer $b_i$ and one seller $s_i$. Each transmission line has capacity 100 and susceptance 0.2.
	Table \ref{tab:example_inf_R} shows, for each seller, the minimum production level, as well as the maximum quantity and the cost of each bid. Also, for each buyer, the price-inelastic demand, the maximum quantity and the valuation of each bid are reported in the same table. We note that only the seller located in node $V_1$ has non-convex preferences.
\end{example}
\begin{table}[H]
	\centering
	\begin{tabular}{ccc|ccc|cccccccccc}
		& $V_1$ & & & $V_2$ & & & $V_3$ \\ 
		$\underline{P}^s_{1}$ & $q^s_{11}$ & $c_{11}$ & $\underline{P}^s_{2}$ & $q^s_{21}$ & $c_{21}$ & $\underline{P}^s_{3}$ & $q^s_{31}$ & $c_{31}$ & \\ \hline
		1000 & 1000 & 30 & 0 & 100 & 15 & 0 & 100 & 25 &  \\ 
	\end{tabular}
	
	\vspace{0.5cm}
	
	\begin{tabular}{ccc|ccc|ccc}
		& $V_1$ & & & $V_2$ & & & $V_3$ \\ 
		$\underline{P}^b_{1}$ & $q^b_{1}$ & $v_{1}$ & $\underline{P}^b_{2}$ &  $q^b_{2}$ & $v_{2}$ & $\underline{P}^b_{3}$ & $q^b_{3}$ & $v_3$ \\ \hline
		40 & 5 & 20 & 25 & 5 & 20 & 20 & 5 & 20 \\ 
	\end{tabular}
	\caption{Example \ref{ex:infeasible_CSWMP} - Parameters}
	\label{tab:example_inf_R}
\end{table}
In this example, $R=\frac{1000}{2}= 500$. Thus, in the perturbed economy, 500 MW of auctioneer demand must be fulfilled \textit{in each node}, summing up to a total of 1500 MW of fictitious demand. However, the total available supply is 1200 MW. Therefore, regardless of the choice of $\alpha$, the convexified scaled welfare maximization problem is infeasible. In other words, a pseudoequilibrium does not exist and no rounding step can be performed to obtain an approximate equilibrium of the original economy. 
Thus, while the choice of $R$ provides a theoretical guarantee, there might not be enough supply on all nodes to guarantee feasibility. 
In fact, in the example at hand, the LP relaxation of the welfare maximization problem (without auctioneer demand) already yields a feasible solution, in which only the seller in $V_2$ is committed and produces at full capacity. 
Even when a pseudoequilibrium exists, $R$ might be chosen too high and result in large inefficiencies of the approximate equilibrium. 
The auctioneer demand was introduced to make sure that there is always sufficient supply to satisfy the demand, but it is hard to estimate. In our implementation, we instead resolve the problem with fixed integer variables (a.k.a. residual clearing) to seek a feasible solution without having to determine auctioneer demand. This is the case, because the quantity produced is determined via continuous variables and only the commitment variables are binary.

\subsection{Implementation for power markets}\label{sec:implementation}

The markup mechanism is based on the assumption that the non-convexities can be thoroughly characterized i.e., that the market operator has access to the concave (convex) hulls of buyers' valuation (sellers' cost) functions as well as to the non-convexity measure $R$. This may not be the case for real-world electricity markets. For example, day-ahead electricity markets are multi-period markets, and buyers or sellers exhibit intertemporal constraints such as ramping or minimum uptime constraints for electricity generators. For such complex cost functions, no explicit representations of convex hulls are readily available, and obtaining convex hulls is computationally expensive \citep{Hua.2017, Schiro.2016}. Similarly, computing the inner radius and inner distance of the non-convex set of feasible solutions might not be practical. Based on the discussion above, we introduce the following modifications for electricity markets:

\begin{itemize}
    \item allocations are identified using a constructive rounding approach, and
    \item no auctioneer demand is included.
\end{itemize}

In what follows, we describe the implementation of the markup mechanism for power markets. This includes a constructive rounding method and adaptations in the search for $\alpha$. 

\subsubsection*{Phase 1 - Convexifying WMP}

We extend the notation introduced in Section \ref{sec:preliminaries} and denote by $z = \left( (x_b)_{b \in B}, (y_s)_{s \in S}, (u_s)_{s \in S}, (f_r)_{r \in R} \right)$ a solution to the welfare maximization problem (WMP)  (\ref{opt:welfare}) associated with a non-convex economy $\mathcal{E}$. 
The variables $(u_s)_{s \in S}$ are binary and encode the sellers' commitment decisions as in \hyperref[opt:dcopf]{DCOPF-MILP}. 
Given an electricity market with transmission network, the scaled welfare for an allocation $z$ and a scale factor $\alpha \geq 0$ is defined as
\begin{align}
    W(z, \alpha) = \sum_{b \in B} \frac{v_b(x_b)}{1+\alpha} - \sum_{s \in S} c_s(y_s) - \sum_{r \in R} d_r(f_r). 
\end{align}
In the first phase, the \textit{convexified scaled welfare maximization problem} is considered:
\begin{align*}
    \label{eq:CSWMP}
    \max_{x, y, u, f} & \ \ W(z, \alpha) \eqname{CSWMP}\\
    \textnormal{ subject to } & z \in \hat{\Omega} \ (\Omega) \\
 & 0 \leq u_s \leq 1 \ \forall \ s \in S. 
\end{align*} 

\noindent
\ref{eq:CSWMP} is the relaxed welfare maximization problem associated with a perturbed economy in which the buyers valuations are scaled down by $(1 + \alpha)$. Also, the valuation (cost) functions from the original economy $\mathcal{E}$ are replaced by their concave (convex) hulls. 
Note that both weak or strict supply can be enforced in \ref{eq:CSWMP}. We comment more on this issue in Section \ref{sec:experiments}. 

For a given $\alpha$, we denote by $z^{\alpha} = \left((x^{\alpha}_b)_{b \in B}, (y^{\alpha}_s)_{s \in S}, (u^{\alpha}_s)_{s \in S}, (f^{\alpha}_r)_{r \in R} \right)$ the optimal solution to \ref{eq:CSWMP} and by $p^{\alpha}$ the associated shadow price vector. Then, $(z^{\alpha}, p^{\alpha})$ is a pseudoequilibrium of the perturbed economy.
Since $z^{\alpha}$ might be infeasible for the WMP, in the second phase of our implementation we round $z^{\alpha}$ to a feasible allocation $z'$ by solving a residual clearing problem (RC) which considers the scaled welfare. We propose two rounding techniques that we describe in the following.

\subsubsection*{Phase 2 - Rounding \& Residual Clearing}

The first method, which we call \textbf{rounding with a threshold} $\delta$, rounds $u^{\alpha}_s$ to 1 if $u^{\alpha}_s \geq \delta$ and to 0 otherwise. Given the rounded $(u^{\alpha}_s)_{s \in S}$ values, the following residual clearing problem is solved:
\begin{align*}
    \label{eq:RC-delta}
    \max_{x, y, u, f} & \ \ W(z, \alpha) \eqname{RC-$\delta$} \\
    \textnormal{ subject to } & z \in \hat{\Omega} \ (\Omega) \\
 & u_s = u^{\alpha}_s \ \forall \ s \in S. 
\end{align*}
The purpose of RC$-\delta$ is two-fold. First, by solving RC$-\delta$ we can verify if the rounded  $(u^{\alpha}_s)_{s \in S}$ values satisfy the feasibility of the WMP. Second, the solution to RC$-\delta$ is the final approximate allocation for the original economy. Given the fixed (rounded) commitment decisions, the initial production/consumption behavior (i.e., the solution computed in the first phase) of the sellers/buyers can be improved in the residual clearing problem. 

Rounding with a threshold can be done in polynomial time. 
However, the welfare of the final allocation varies depending on the choice of $\delta$, since a small (large) value for $\delta$ may lead to overcommiting (undercommiting) sellers. Therefore, a search routine\footnote{To optimize computation time, the search routine can be parallelized i.e., multiple instances of \ref{eq:RC-delta} with different values for $\delta$ are solved in independent threads.} is required to find the value of $\delta$ that leads to a feasible solution with the highest welfare. Moreover, we remark that with maximum uptime constraints added, such a \emph{``greedy''} approach ceases to work. This motivates consideration of an alternative rounding method.

The second technique that we propose is \textbf{rounding by solving a small MILP}. Let $S^{\alpha}$ be the set of sellers such that $u^{\alpha}_s \in \{0, 1\}$ if $s \in S^{\alpha}$,
where $u^{\alpha}_s$ is part of the pseudoequilibrium allocation computed in the first phase. A solution to the WMP can be computed by solving the following non-convex residual clearing problem:
\begin{align*}
    \label{eq:RC-MILP}
    \max_{x, y, u, f} & \ \ W(z, \alpha) \eqname{RC-MILP} \\
    \textnormal{ subject to } & z \in \hat{\Omega} \ (\Omega) \\
 & u_s  = u^{\alpha}_s \ \forall \ s \in S^{\alpha} \\
 & u_s \in \{0,1\} \ \forall \ s \in S \setminus S^{\alpha}. 
\end{align*} 
Note that the rounding routine is integrated in the residual clearing problem \ref{eq:RC-MILP}. While \ref{eq:RC-MILP} integer programs of this sort cannot be solved in polynomial time, the computational cost are proportional to the size of $S^{\alpha}$. For smaller problems, the number of fractional variables $u_s$ can be such that the MILP-based rounding can be solved in due time.%, and significantly faster than MWP.

Rounding a fractional solution to a feasible integer solution is NP-hard in general. However, in our experiments with realistic data sets in Section \ref{sec:experiments} we show that \ref{eq:RC-delta} and \ref{eq:RC-MILP} often compute feasible solutions providing a practical approach with substantial runtime benefits over an exact solution to the WMP for large instances.

\subsubsection*{Finding a good $\alpha$} In phases 1 and 2 we assumed that $\alpha$ is given. Obviously, the quality of the approximate equilibrium depends on the choice of $\alpha$. \citet{Milgrom.2021} propose to identify a value for $\alpha$ such that the prices for the sellers $p^{\alpha}$ and the prices for the buyers $(1 + \alpha)p^{\alpha}$ which are associated to the pseudoequilibrium of the perturbed economy guarantee no budget deficit in $\mathcal{E}$. 
As we discussed earlier, we have a very large number of goods such that $R$ might lead to infeasibilities. Without auctioneer demand of this sort, we cannot guarantee that every participant gets an element of their demand set and some will require MWPs. 

Given these considerations, we aim to find the smallest $\alpha$ from a set of predefined values, such that the market incurs no budget deficit after paying the MWPs with buyers' money. Concretely, given a set $I$, we select the smallest $\alpha \in I$ such that 

\begin{equation}
    p^T \left( \sum_{b \in B} x_b (1+\alpha) - \sum_{s \in S} y_s -  \sum_{r \in R} f_r \right) - \text{MWPs} \geq 0.
\end{equation}
To avoid large differences between the buyers' and the sellers' payments, we only consider values for $\alpha$ that are smaller than 1. Note that, depending on $I$, there might not exist $\alpha \in I$ that satisfies this condition. 

\section{Experimental evaluation} \label{sec:experiments}

In the following, we summarize the results of our experimental evaluation. First, we describe the data sets used in our study before we outline the results for smaller and large data sets. For the smaller data sets with less than 11k nodes, we can compute the exact solution of the underlying welfare maximization problem (\hyperref[opt:dcopf]{DCOPF-MILP}) in less than one hour, a time limit that is still relevant for the day-ahead market. For the EU electricity market the time limit is only 17 minutes.

\subsection{Data sets}

Our results are based on the data sets for the ARPA-E Grid Optimization Competition \citep{arpa-competition} which provides several realistic power system descriptions and operating scenarios.  
We evaluated the markup mechanism on six scenarios from the ARPA-E GO Competition Challenge 2\footnote{\url{https://gocompetition.energy.gov/challenges/challenge-2}}, which are described in Table \ref{tab:datasets}.
Each scenario covers one period and includes network, load and generating data. 
We also computed the optimal, welfare maximizing solutions and supporting IP prices, following the approach that is currently employed in the U.S. ISO markets.
As discussed in Section \ref{sec:preliminaries_allocation_pricing}, the EU market clearing algorithm computes zonal prices and the bid language differs from that in U.S. markets. 

\begin{table}[!ht]
    \centering
    \begin{tabular}{|r|r|r|r|r|r|r|r|}
    \hline
        \textbf{Nodes} & \textbf{Lines} & \textbf{Sellers} & \textbf{Buyers} & \textbf{Demand} & \textbf{Supply} & \textbf{Variables} & \textbf{Constraints} \\ \hline
        617 & 841 & 94 & 404 & 506,361.82 & 830,928.96 & 109,752 & 244,413 \\ \hline
        2,020 & 2,709 & 194 & 1,392 & 622,224.37 & 1,041,339.21 & 344,536 & 770,141 \\ \hline
        8,718 & 26,084 & 321 & 5,819 & 988,735.25 & 1,088,344.65 & 1,447,800 & 3,311,227 \\ \hline
        10,480 & 16,253 & 750 & 6,871 & 2,033,795.10 & 2,289,028.45 & 1,817,280 & 4,160,463 \\ \hline
        12,209 & 30,500 & 2,009 & 6,987 & 2,391,960.62 & 3,328,699.92 & 2,033,280 & 4,548,179 \\ \hline
        18,889 & 57,630 & 898 & 12,273 & 2,686,818.18 & 2,916,900.60 & 3,165,528 & 7,279,122 \\ \hline
    \end{tabular}
    \caption{Characteristics of 6 data sets from the ARPA-E Grid Optimization Challenge 2, extended to 24 hours datasets. Supply and demand are measured in MWh. The last two columns contain the number of variables and constraints of the welfare maximization problem.}
    \label{tab:datasets}
\end{table}

Note that the ARPA-E data sets model a single period (hour) problem only. A key difference to clearing problems on power markets is the fact that demand and supply bids span multiple periods. Generators exhibit uptime constraints and the demand-side often requires contiguous amounts of power for multiple hours. 
To evaluate the markup mechanism on realistic day-ahead markets, we extend the original ARPA-E datasets to scenarios with 24 hours as they need to be solved on day-ahead markets. The data for these scenarios is available online\footnote{\url{https://gitlab.lrz.de/ge72yeq/markup-mechanism-for-electricity-markets/-/tree/main/data?ref_type=heads}}.

On the supply side, we first partition the set of generators into renewable (RES) and non-renewable (Non-RES) resources. We perform this split because the maximum production of renewable resources varies over the day and we want to model supply fluctuations accordingly. Thus, we partition the generators based on the no-load cost and assume that any resource with zero no-load cost belongs to RES. 
All other resources belong to Non-RES. We assume that there are two types of RES resources, namely wind and photovoltaic units. For each resource in RES we choose one of these types with equal probability.

Next, we introduce supply fluctuations following the percentages for the hourly variation of wind and solar units that are provided in the data for the EU bidding zone review (BZR)\footnote{Tabs "Wind" and "Solar" in the input files available here: \url{https://www.entsoe.eu/network_codes/bzr/}}. 
We assume that the original ARPA-E data corresponds to February 01, 2009, 7am-8am and is modeled as the 8th period in the extended scenario. Hence, for each RES unit, the maximum production corresponding to the 8th period is identical to the maximum production provided in the original ARPA-E dataset.
We compute the maximum production from the remaining periods by multiplying the corresponding hourly percentage factor from the BZR data by the maximum production specified in the ARPA-E data.
Since the Non-RES resources are assumed to be dispatchable, we do not alter their maximum production over different periods. 

We assume that resources in RES have no minimum uptime, while resources in Non-RES are assigned a minimum uptime depending on their maximum production. Thus, with equal probability, we set a minimum uptime of 0 (i.e., no minimum uptime), 4 or 6 hours for the generators that have a maximum production smaller than 1500MW/h, and no minimum uptime for the generators that have maximum production larger than 1500MW/h.

\subsection{Results}

For our computations, we draw on the Gurobi optimizer 10.0 \citep{Gurobi.2023}. The experiments were conducted on a computer with Intel Core Processor i7-1260P with 12 cores and 32 GB RAM running on a Microsoft Windows 11 operating system. 
We also ran our experiments for larger instances on a powerful Linux cluster of 96 nodes with 56 GB RAM on each node (28-way Haswell-based nodes and FDR14 Infiniband interconnect). We find that the additional computing power does not change the results that we can achieve within one hour. For the test instances with more than 18,889 nodes, OPT can (at most) be solved with 24-hour, but such time limits are irrelevant for day-ahead markets. For the one hour time limit, we are interested in, the compute cluster did not lead to much better results. In what follows, we report the results based on the Intel Core Processor i7-1260P with 12 cores.

We analyzed the experimental results for all six data sets, which are summarized in Table \ref{tab:results}.
In the remainder of this section we denote by OPT the optimal, welfare maximizing allocation (or, depending on the context, its associated welfare).
For the scenarios with less than 11k nodes, we provide the results for the welfare maximizing mixed integer linear program (\hyperref[opt:dcopf]{DCOPF-MILP}) with exact supply-demand equivalence, which we compare to the results of the markup mechanism. Apart from a solution with exact supply-demand equivalence in the markup mechanism, we also report solutions where we allow for modest levels of excess supply, which can sometimes be necessary for a feasible solution. If the day-ahead market has excess supply, the system operator could provide this excess supply on the intraday or real-time market. Note that the original markup mechanism deals with allocation problems that do not require a strict demand-supply equivalence. 

The sizes of the resulting MILPs vary depending on the scenario. For instance, the MILP corresponding to the 617-node instance includes 107,496 continuous and 2,256 binary variables, while the MILP associated to the 10,480-node scenario has 1,799,280 continuous and 18,000 binary variables.
Table \ref{tab:results} also reports the welfare gains achieved, the supply and demand cleared as well as the level of oversupply in those cases where we do not enforce strict demand-supply equivalence. We summarize our main result below.

\begin{result}
    The markup mechanism computes a close approximation of OPT ($<0.01\%$ welfare loss). The runtime of the markup mechanism is significantly smaller than the runtime required to compute OPT, for all instances where OPT could be computed within one hour. For larger instances beyond 12 nodes, neither OPT nor MW could be solved within two hours.
\end{result}

For all scenarios that we consider, the markup mechanism finds allocation solutions with RWL at most 0.04\%, but typically 0.01\% or less. We conjecture that this low welfare loss is due to the fact that the problems have a large number of continuous variables compared to the integer unit commitment variables. 
Furthermore, we notice that the difference between the runtime required to compute OPT and the runtime of the markup mechanism becomes larger as we consider more larger scenarios. For the instance with 617 nodes, threshold-based rounding with $\delta = 0.5, \alpha = 0.01$ has runtime equal 4.75s, while computing OPT takes 138.06s. The same rounding technique has runtime equal 333.49s ($\sim$6 minutes) for the scenario with 8,718 nodes, while calculating the corresponding optimal solution requires 3,108.48s ($\sim$52 minutes). For the cases with 12,209 and 18,889 nodes, respectively, OPT could not be computed within the one hour time limit that we set. 

MWPs are required for IP Pricing, if we can compute the optimal integer solution, but they might also be required in the markup mechanism for electricity markets. With IP Pricing and an optimal solution, the total MWPs constitute the budget deficit. In the markup mechanism, we can accumulate a budget surplus, but still have MWPs, as we discussed earlier. In Table \ref{tab:results} we report the MWPs separately, and a budget deficit in brackets describes a surplus \textit{after} deducting the costs for the MWPs. 
Thus, besides its approximation and computational properties, the markup mechanism achieves good results in terms of budget feasibility and individual rationality.

\begin{result}
    The markup mechanism leads to no budget deficit. Moreover, the MWPs required in the markup mechanism are smaller than the MWPs achieved with IP Pricing applied on OPT.
\end{result}

\newcolumntype{R}[1]{>{\raggedleft\let\newline\\\arraybackslash\hspace{0pt}}m{#1}}
\newcommand*{\thead}[1]{\multicolumn{1}{|c|}{\bfseries #1}}

% \newpage
\begin{sidewaystable}
    \centering
    \resizebox{\columnwidth}{!}{
       \begin{tabular}{|l|r|r|r|r|r|r|r|r|r|r|r|}
        \hline
        \thead{\textbf{Algorithm}} & \thead{\textbf{S=D}} & \thead{\textbf{Welfare}} & \thead{\textbf{Supply}} & \thead{\textbf{Demand}} & \thead{\textbf{Oversupply}} & \thead{\textbf{MWPs}} & \centering\textbf{Budget} \newline \textbf{ Deficit} & \thead{$\boldsymbol{\alpha}$} & \thead{$\boldsymbol{\delta}$}  & \thead{\textbf{RWL}} & \thead{\textbf{Time (s)}} \\ \hline
				\textbf{617 nodes} &&&&&&&&&&&\\ \hline
        OPT & y & 238,195,146.54 & 298,663.25 & 298,663.25 & 0 & 86,336.04 & 86,336.04 & ~ & ~ & ~ & 138.06 \\ \hline
        Threshold & y & 238,174,692.20 & 299,450.34 & 299,450.34 & 0 & 47,262.63 & 47,262.63 & 0 & 0.01	& 0.01\% & 2.64  \\ \hline
        Threshold & y & 238,104,566.44 & 289,878.27 & 289,878.27 & 0 & 4,519.08	& (190,312.38) & 0.01 &	0.5	& 0.04\% & 4.75 \\ \hline
        Threshold & y & 238,075,360.52 & 284,765.32 & 284,765.32 & 0 & 4,914.59 & (1,781,451.71) & 0.1 & 0.5 & 0.04\% & 2.45 \\ \hline
        Threshold & n & 238,193,906.57 & 323,214.68 & 299,853.02 & 23,361.65 & 27,732.20 & (172,037.90)	& 0.01 & 0.2 & 0.00\% & 3.28 \\ \hline
		MILP Round & n & 238,197,997.90 & 322,965.59 & 299,410.14 & 23,555.45 & 16,591.57 & (182,998.77) & 0.01 & &	0.00\% & 73.92 \\ \hline	
                \textbf{2,020 nodes} &&&&&&&&&&&\\ \hline
        OPT & y & 597,582,514.99 & 284,313.42 & 284,313.42 & 0 & 149,029.65 & 149,029.65 & ~ & ~ & ~ & 795.28 \\ \hline
        Threshold & y & infeasible &  &  &  &  &  & 0.1 & 0.5 & & \\ \hline
        Threshold & y & 597,567,977.75 & 285,761.50 & 285,761.50 & 0 & 71,262.31 & (34,085.29) & 0.01 & 0.1 & 0.00\% & 11.09 \\ \hline
        Threshold & y & 597,575,055.59 & 284,209.02 & 284,209.02 & 0 & 66,401.13 & (957,183.18) & 0.1 & 0.1 & 0.00\% & 23.13 \\ \hline
        Threshold & n & 597,942,187.64 & 336,626.35	& 288,411.20 & 48,215.14 & 50,253.30 & (34,237.35) & 0.01 & 0.10 & -0.06\% & 32.32 \\ \hline        
        MILP Round & n & 597,930,887.70	& 319,175.01 & 287,584.56 & 31,590.45 & 56,555.09 & (761,305.48) & 0.10 & & -0.06\% & 121.60 \\ \hline    
                \textbf{8,718 nodes} &&&&&&&&&&&\\ \hline
        OPT & y & 53,762,905.20 & 803,462.34 & 803,462.34 &	0 &	11,851.68 & 11,851.68 & ~ & ~ & ~ & 3,108.48 \\ \hline
        Threshold & y & 53,762,321.96 & 803,165.67 & 803,165.67 & 0 & 48.61 & (310,681.06) & 0.01 & 0.5 & 0.00\% & 333.49 \\ \hline
				\textbf{10,480 nodes} &&&&&&&&&&&\\ \hline
        OPT & y & 82,630,373.56	& 1,610,608.60 & 1610608.597 & 0 & 56,396.69 & 56,396.69 & & & & 14,454.78 \\ \hline
        Threshold & y & 82,628,972.82 &	1,609,101.04 & 1,609,101.04 & 0 & 5,089.26 & (512,300.75) & 0.01 & 0.5 & $0.00\%$ & 2,271.02 \\ \hline
            \textbf{12,209 nodes} &&&&&&&&&&&\\ \hline
        OPT & y & N/A &  &  &  &  &  & & & &  \\ \hline
        Threshold & y & 901,291,505.82 & 1,227,340.33 & 1,227,340.33 & 0 & 237,410.14 & (352,738.44) & 0.01 & 0.01 & - & 1,445.82 \\ \hline
            \textbf{18,889 nodes} &&&&&&&&&&&\\ \hline
        OPT & y & N/A	&  &  &  &  &  & & & &  \\ \hline
        Threshold & y & 54,586,238.75 & 2,081,555.98 & 2,081,555.98 & 0 & 20,477.98 & (377,467.48) & 0.01 & 0.01 & - & 49,178.68 \\ \hline
    \end{tabular}}
    \caption{Computed outcomes of the markup mechanism for the 617, 2020, 8718, 10,480, 12,209 and 18,889 node instances from the ARPA-E Grid Optimization Challenge 2, extended to 24 hour datasets. The second column denotes whether supply-demand equivalence is strict or weak. Welfare, MWPs and the budget deficit are measured \euro, where for the latter a quantity in parentheses denotes a surplus. Supply, demand and oversupply, in turn, are measured in MWh.}
    \label{tab:results}
\end{sidewaystable}

With $\alpha > 0$ we achieve budget surplus in all four scenarios, regardless of the rounding technique that is employed. Also, as $\alpha$ increases, the budget surplus becomes larger, as we notice in e.g. the scenario with 617 nodes, in which a budget surplus of \euro190,312.38 is achieved with $\alpha = 0.01$, while $\alpha = 0.1$ leads to extra payments in the amount of \euro1,781,451.71 (threshold-based rounding with $\delta = 0.5$). 
For the scenario with 2,020 nodes, the MWPs seem to be negatively correlated with $\alpha$ (e.g., with threshold-based rounding and $\delta = 0.1$, MWPs = 71,262.31 for $\alpha = 0.01$ and MWPs = 66,401.13 for $\alpha = 0.1$). However, MWPs do not depend on $\alpha$ monotonically in general.

As described in Section \ref{sec:mw}, the threshold-based rounding technique relies on the hyperparameter $\delta$. In particular, the optimal value for $\delta$ is scenario-specific. For the instance with 617 nodes, $\delta = 0.5$ led to a solution having the smallest MWPs among all $(\alpha, \delta)$ configurations that we considered (MWPs=4,519.08 with $\alpha = 0.01$), while for the scenario with 2,020 nodes the same $\delta$-value caused the infeasibility of the second phase, for any $\alpha \in \{0, 0.01, 0.1\}$. Furthermore, the experimental results show that the MWPs decrease as $\delta$ increases, as less generators are committed if a large $\delta$-value is employed. Yet, the monotonicity of MWPs depending on $\delta$ is, in general, not guaranteed. 

Enforcing weak supply-demand equivalence in the markup mechanism and thus allowing for modest levels of oversupply is an approach that increases the welfare, as the results for the scenarios with 617 and 2,020 nodes indicate. To compute the results corresponding to the 617- and 2,020-nodes scenarios with weak supply-demand equivalence, we included an additional constraint in both phases of the markup mechanism, that requires the oversupply to be at most half the fulfilled demand in the current allocation. 
We note that for the scenario with 617 nodes, MILP rounding computes a solution having oversupply equal to $\sim$7\% of the total supply. For the 2,020-nodes scenario the same rounding technique finds a solution having oversupply equaling $\sim$10\% of the total supply. Both the threshold-based and the MILP rounding methods compute outcomes with lower MWPs compared to the best solution computed by the markup mechanism with strict supply-demand equivalence.

\section{Conclusions} \label{sec:conclusion}

Electricity spot markets in the U.S. and in the EU are growing. Due to renewable energy sources, there will be more market participants, leading to growing problem instances for the allocation and pricing problems that market operators need to solve. \citet{Milgrom.2021} introduced the markup mechanism for computing allocation and prices in non-convex markets. The markup solution can be computed with only convex optimization algorithms, and the mechanism is nearly-efficient and has no budget deficit. 
Electricity markets are a prime candidate for such techniques. While MILP techniques have become so powerful that large problems can be solved to optimality in practice, these techniques find their limits on today's electricity markets. 
However, the application of the mechanism requires appropriate rounding techniques. The original approach relies on existence theorems such as those by Shapley and Folkman. Rounding is hard in general, but for many electricity market problems simple threshold-based rounding works well, as we show in this paper. 
However, with simple rounding, individual rationality of the solution cannot be guaranteed and some of the market participants require make-whole payments to break even. We introduce a two-stage polynomial-time solution process to deal with these problems. 

In addition, we contribute a thorough empirical evaluation of the approach based on realistic data from the ARPA-E Grid Optimization Competition. For problem sizes up to 10,480 nodes our implementation of the markup mechanism computes solutions substantially faster than exact MILP techniques. For an instance with 12,209 nodes, the optimal solution could not be computed anymore within several hours, but the markup solution was obtained within 25 minutes. Larger problems (18,000 nodes or more) could not be solved with either approach within 12 hours even on a compute cluster. The computational speedups of the markup mechanism are relevant for practice. The welfare loss is surprisingly low and so is the markup for buy-side prices (usually 1\%). Apart from the computational speedup, the markets do not lead to a budget deficit and the make-whole payments are substantially lower than those computed with standard IP pricing. As such, the markup mechanism by \citet{Milgrom.2021} might well provide an alternative for future electricity markets.

\subsection*{Acknowledgments}

This work was supported by the German Research Foundation (Deutsche Forschungsgemeinschaft, DFG) [BI 1057/10].

\bibliographystyle{informs2014} 
\bibliography{bibliography} 

\begin{appendices}

    \section{DCOPF Problem} \label{app:dcopf}

\begin{table}[H]\centering
       \def\sym#1{\ifmmode^{#1}\else\(^{#1}\)\fi}
       \begin{tabular}{l*{2}{l}}
       \toprule
              \textbf{Input}      &     &     \textbf{Description}  \\
                           % &\multicolumn{1}{c}{(1)}     &\multicolumn{1}{c}{(3)}   \\
       \midrule
       $B$            &   & Set of buyers \\
       $S$            &   & Set of sellers \\
       $V$            &   & Set of network nodes \\
       $E$ & & Set of network lines \\
       $T$            &   &  Set of time periods (hours)\\
       $R^*$            &   & Reference node \\
       $N(v)$            &   & Node-neighbors of node $v$ (both out- and in-going nodes) \\
       $\upsilon(i)$  & $B \cup S \rightarrow V $ &  Node in which agent $i \in B \cup S$ is located\\
       $\beta_i^t$            &   & Bids of agent $i \in B \cup S$ in period $t \in T$ \\
       $v_{btl}$            &  [\euro/MWh] & Value of bid $l \in \beta_b^t$ of buyer $b \in B$ in period $t\in T$ \\
       $q_{btl}$            &  [MWh] & Max. quantity of bid $l \in \beta_b^t$ of buyer $b \in B$ in period $t\in T$ \\
       $c_{stl}$            &  [\euro/MWh] & Marginal cost of bid $l \in \beta_s^t$ of seller $s \in S$ in period $t\in T$ \\
       $h_{st}$            &  [\euro] & No-load cost of seller $s \in S$ in period $t \in T$ \\
       $q_{stl}$            & [MWh] & Max. quantity of bid $l \in \beta_s^t$ of seller $s \in S$ in period $t\in T$ \\
       $\underline{R}_s$ & & Min. uptime of seller $s \in S$ (hours) \\
       $\underline{P}_{bt}$ & [MWh] & Price-inelastic demand of buyer $b \in B$ in period $t\in T$ \\
       $\overline{P}_{bt}$ & [MWh] & Max. demand of buyer $b \in B$ in period $t\in T$ \\
       $\underline{P}_{st}, \overline{P_{st}}$ & [MWh] & Min. and max. output of seller $s \in S$ in period $t\in T$ \\
       $B_{vw}$ & [pu] & Susceptance of the line from $v \in V$ to $w \in V$\\
       $\underline{F}_{vw}, \overline{F}_{vw}$ & & Min. and max. allowed flow on the line from $v \in V$ to $w \in V$\\
       \midrule
       \end{tabular}
       \caption{DCOPF Input Parameters}
       \label{dcopf-input-params}
\end{table}

\begin{table}[H]\centering
       \def\sym#1{\ifmmode^{#1}\else\(^{#1}\)\fi}
       \begin{tabular}{l*{2}{l}}
       \toprule
              \textbf{Variable}      &     &     \textbf{Description}  \\
                           % &\multicolumn{1}{c}{(1)}     &\multicolumn{1}{c}{(3)}   \\
       \midrule
       $x_{btl} \geq 0$ & [MW] & Consumption from bid $l \in \beta_b^t$ for buyer $b \in B$ and $ t\in T$   \\
       $y_{stl} \geq 0$ & [MW] & Production from bid $l \in \beta_s^t$ for seller $s \in S$ and $ t\in T$   \\
       $x_{bt} \geq 0$ & [MW] & Total consumption of buyer $b \in B$ in period $t\in T$ \\
       $y_{st} \geq 0$ & [MW] & Total production of seller $s \in S$ in period $t\in T$\\
       $u_{st} \in \{0, 1\}$ & & Commitment of seller $s \in S$ in period $t\in T$\\
       $\phi_{st} \geq 0$ & & Startup indicator of seller $s \in S$ in period $t\in T$\\
       $\alpha_{vt} \in \mathbb{R}$ & [rad] & Voltage angle at node $v \in V$ in period $t \in T$\\
       $f_{vwt} \in \mathbb{R}$ & [MW] & Flow on the line between $v, w \in V$ in period $t \in T$\\
       \midrule
       \end{tabular}
       \caption{DCOPF Variables}
       \label{dcopf-primal-var-description}
\end{table}

\newpage

\begin{align}
       \max\limits_{x, y, u, f, \alpha}  & \displaystyle\sum_{b \in B} \sum_{t \in T} \sum_{l \in \beta_b^t} v_{btl} x_{btl} - 
       \sum_{s \in S} \sum_{t \in T} \sum_{l \in \beta_s^t} c_{stl} y_{stl} - \sum_{s \in S} \sum_{t \in T} h_s u_{st} \label{opt:dcopf} & \tag{DCOPF-MILP} \\
       \text{subject to}& \displaystyle \sum_{s \in S | \nu(s) = v}y_{st} - \sum_{b \in B | \nu(b) = v} x_{bt} - \sum_{w \in N(v)}f_{wvt} = 0 & \forall v \in V, t \in T \nonumber\\
       &  \displaystyle f_{vwt} - B_{vw}(\alpha_{vt} - \alpha_{wt}) = 0 & \forall v \in V, w \in N(v), t \in T \nonumber\\
       & f_{vwt} \leq \overline{F}_{vw} & \forall v \in V, w \in N(v), t \in T \nonumber\\
       & f_{vwt} \geq \underline{F}_{vw} & \forall v \in V, w \in N(v), t \in T \nonumber\\
       & x_{btl} \leq q_{btl} & \forall b \in B, t \in T, l \in \beta_b^t \nonumber\\
       & x_{bt} - \sum\limits_{l \in \beta_b^t} x_{btl} = \underline{P}_{bt} & \forall b \in B, t \in T \nonumber\\
       & x_{bt} \leq \overline{P}_{bt} & \forall b \in B, t \in T \nonumber\\
       & y_{stl} \leq q_{stl} u_{st} & \forall s \in S, t \in T, l \in \beta_s^t \nonumber\\
       & y_{st} - \sum\limits_{l \in \beta_s^t} y_{stl} = 0 & \forall s \in S, t \in T \nonumber\\
       & y_{st} - \overline{P}_{st} u_{st} \leq 0 & \forall s \in S, t \in T \nonumber\\
       & y_{st} - \underline{P}_{st} u_{st} \geq 0 & \forall s \in S, t \in T \nonumber\\
       & \sum\limits_{i=t-\underline{R}_s + 1}^{t} \phi_{si} - u_{st} \leq 0 & \forall s \in S, 1<t\leq T \nonumber\\
       & \phi_{st} - u_{st} + u_{s(t-1)} \geq 0 & \forall s \in S, 1<t\leq T \nonumber\\
       & \alpha_{R^*t} = 0 & \forall t \in T \nonumber
       \end{align}

% \begin{table}[H]\centering
% \def\sym#1{\ifmmode^{#1}\else\(^{#1}\)\fi}
% % \caption[Short Heading]{\protect\lipsum[1]}
% \begin{tabular}{l*{2}{l}}
% \toprule
%        \textbf{Dual variable}      &     &     \textbf{Description}  \\

% \midrule
% $p_{vt} \in \mathbb{R}$ & [\euro/MWh] & Price at node $v \in V$ in period $t \in T$\\
% $\gamma_{vwt} \in \mathbb{R}$ & [\euro/MWh] & Congestion price for the line between $v, w \in V$ in \\
% & & period $t \in T$\\
% $r_t \in \mathbb{R}$ & & Dual of the reference node voltage angle constraint in \\ 
% & & period $t \in T$ \\
% \midrule
% \end{tabular}
% \caption{DCOPF Dual variables}
% \label{dcopf-dual-var-description}
% \end{table}

\end{appendices}

\end{document}